\documentclass[twocolumn,showpacs,preprintnumbers,amsmath,amssymb,superscriptaddress]{revtex4}

\usepackage{graphicx}
\begin{document}
\bibliographystyle{prsty}
\title{Chemical potential landscape in band filling and bandwidth-control of manganites: Photoemission spectroscopy measurements}

\author{K. Ebata}
\affiliation{Department of Physics and Department of Complexity Science and Engineering, University of Tokyo, 7-3-1 Hongo, Bunkyo-ku, Tokyo 113-0033, Japan}
\author{M. Takizawa}
\affiliation{Department of Physics and Department of Complexity Science and Engineering, University of Tokyo, 7-3-1 Hongo, Bunkyo-ku, Tokyo 113-0033, Japan}
\author{A. Fujimori}
\affiliation{Department of Physics and Department of Complexity Science and Engineering, University of Tokyo, 7-3-1 Hongo, Bunkyo-ku, Tokyo 113-0033, Japan}
\author{H. Kuwahara}
\affiliation{Department of Physics, Sophia University, Chiyoda-ku, Tokyo 102-8554, Japan}
\author{Y. Tomioka}
\affiliation{Correlated Electron Research Center (CERC), National Institute of Advanced Industrial Science and Technology (AIST), Tsukuba 305-8562, Japan}
\author{Y. Tokura}
\affiliation{Correlated Electron Research Center (CERC), National Institute of Advanced Industrial Science and Technology (AIST), Tsukuba 305-8562, Japan}
\affiliation{Department of Applied Physics, University of Tokyo, Bunkyo-ku, Tokyo 113-8656, Japan}
\affiliation{Spin Superstructure Project, Exploratory Research for Advanced Technology (ERATO), Japan Science and Technology Corporation (JST), Tsukuba 305-8562, Japan}
\date{\today}

\begin{abstract}
We have studied the effects of band filling and bandwidth control on the chemical potential in perovskite manganites $R_{1-x}A_x$MnO$_3$ ($R$ : rare earth, $A$ : alkaline earth) by measurements of core-level photoemission spectra. A suppression of the doping-dependent chemical potential shift was observed in and around the CE-type charge-ordered composition range, indicating that there is charge self-organization such as stripe formation or its fluctuations. As a function of bandwidth, we observed a downward chemical potential shift with increasing bandwidth due to the reduction of the orthorhombic distortion. After subtracting the latter contribution, we found an upward chemical potential shift in the ferromagnetic metallic region $0.3<x<0.5$, which we attribute to the enhancement of double-exchange interaction involving the Jahn-Teller-split $e_g$ band.
\end{abstract}


\maketitle
Perovskite-type manganites with the formula $R_{1-x}A_x$MnO$_3$, where $R$ is a rare-earth and $A$ is an alkaline-earth metal, exhibit a complex phase diagram as a function of band filling and bandwidth, and competition between these phases leads to remarkable phenomena such as colossal magnetoresistance and the spin, charge and orbital ordering of Mn 3$d$ $e_g$ electrons as shown in Fig. 1 \cite{Kuwahara, Tomioka, Tokura5, Kajimoto_band}. Key features to understand the complex phase diagram are double-exchange interaction and the instabilities towards spin, charge and orbital ordering. Systems with small and intermediate bandwidths such as Pr$_{1-x}$Ca$_x$MnO$_3$ (PCMO) and Nd$_{1-x}$Sr$_x$MnO$_3$ (NSMO) exhibit the so-called CE-type antiferromagnetic (AF) charge-ordered (CO) phase in the doping range around half-doping $x$ = 0.5, but this tendency disappears in systems with large bandwidths such as La$_{1-x}$Sr$_x$MnO$_3$ (LSMO). 
This clearly demonstrates the importance of band filling and bandwidth to understand the physical properties of the manganites.

The electron chemical potential $\mu$ is one of the most fundamental physical quantities of strongly correlated electron systems. The shift of $\mu$ as a function of electron density $n$ corresponds to the charge compressibility $\kappa$ or the charge susceptibility $\chi_c$ through $\kappa = (1/n^2)(\partial n/\partial \mu$) or $\chi_c = \partial n/\partial \mu$ and can be measured through the shifts of photoemission spectra since binding energies in the spectra are experimentally referenced to $\mu$, namely, the Fermi level. Since $n$ is determined by the number of electrons in a unit cell and the unit cell volume and the bandwidth are closely related with each other, it is highly important to study the chemical potential shift $\Delta \mu$ as a function of both band filling and bandwidth. Recently, suppression of $\Delta \mu$ as a function of hole doping has been observed in and near the CE-type CO composition range of PCMO and its correlation with the changes of the periodicity of the stripe fluctuations has been pointed out \cite{Ebata}. Pinning of chemical potential by static or dynamic stripe fluctuations in La$_{2-x}$Sr$_x$CuO$_4$ and La$_{2-x}$Sr$_x$NiO$_4$ has also been observed in the composition range where the periodicity of charge order changes with band filling \cite{Ino, Satake}. $\Delta \mu$ in LSMO, which has the widest bandwidth among the manganites, exhibited a monotonous shift without indication of chemical potential pinning, reflecting that there is no intrinsic stripe formation in LSMO \cite{Matsuno, Horiba}. As for the effect of double-exchange interaction on $\Delta \mu$, an upward $\Delta \mu$ with decreasing temperature has been found in the low-temperature part of the ferromagnetic metallic (FM) phase and attributed the shift to the change of bandwidth under Jahn-Teller-split $e_g$ band induced by double-exchange interaction \cite{Ebata3}. This observation is consistent with the theoretical prediction using one-orbital double-exchange model by Furukawa \cite{Furukawa}.

In this paper, we address the issue of the effect of bandwidth control on $\Delta \mu$ through the core-level photoemission measurements of the doping-dependent $\Delta \mu$ of the small bandwidth system PCMO, the intermediate bandwidth system NSMO and the large bandwidth system LSMO and deduce the $\Delta \mu$ as a function of both band filling and bandwidth. In addition to the suppression of the doping-dependent $\Delta \mu$ for hole concentrations near and in the CE-type CO composition range, we observed $\Delta \mu$ as a function of the A-site ionic radius $\langle r_A \rangle$ induced by the orthorhombic distortion and double-exchange interaction.
\begin{figure}
\begin{center}
\includegraphics[width=6cm]{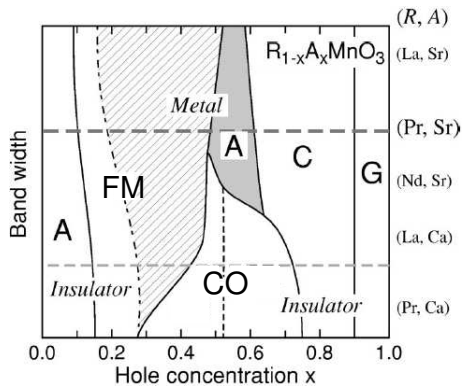}
\caption{Schematic phase diagram of the perovskite-type manganites $R_{1-x}A_x$MnO$_3$ in the band filling (hole concentration)-bandwidth plane at low temperatures \cite{Kajimoto_band}. FM, A, CO, C, and G denote the ferromagnetic metallic, A-type antiferromagnetic, charge-ordering, C-type antiferromagnetic, and G-type antiferromagnetic phases, respectively.}
\label{phase}
\end{center}
\end{figure}

Single crystals of NSMO ($x=0.4$, 0.45, 0.5, 0.55, 0.6, and 0.7), LSMO ($x=0.2$, 0.3, 0.4, 0.45, 0.5, and 0.55) and PCMO ($x=0.2$, 0.25, 0.3, 0.35, 0.4, 0.5, 0.6 and 0.65) were prepared by the floating zone method \cite{Kuwahara, Urushibara, Tomioka}. X-ray photoemission measurements were performed using a Mg $K \alpha$ source ($h\nu =$ 1253.6 eV) and a SCIENTA SES-100 analyzer. The measured binding energies were stable, because the gold $4{\it f}_{7/2}$ core-level spectrum did not change in the measurements with the accuracy of $\pm 10$ meV at each temperature.
All the photoemission measurements were performed under the base pressure of $\sim 10^{-10}$ Torr at 300 K and 80 K. The sample surfaces were repeatedly scraped {\it in situ} with a diamond file to obtain clean surfaces.

Figure 2(a) and (b) shows the spectra of the O $1s$ and Sr $3d$ core levels, respectively. The vertical lines mark the estimated positions of the core levels employed to evaluate the shifts. For the O $1s$ core level, we used the midpoint of the low binding-energy slope because the line shape on the higher binding energy side of the O $1s$ spectra is known to be sensitive to surface contamination or degradation. We also used the midpoint for the Sr $3d$ core level for the same reason. The shifts of the Nd $3d$ and Mn $2p$ core levels (not shown) have been determined from their peak positions because their line shapes slightly changed with composition.

In Fig. 2(c), we have plotted the binding energy shift $\Delta E_B$ of each core level as a function of hole concentration. One can see that the O $1s$, Sr $3d$ and Nd $3d$ core levels are shifted in the same direction while the Mn $2p$ core level are shifted in the opposite direction.
The opposite shift of the Mn $2p$ core level can be explained by the change of the Mn valence with hole doping \cite{Ebata}.
Therefore, we conclude that nearly the same shifts of the O $1s$, Sr $3d$ and Nd $3d$ core levels reflect $\Delta \mu$, and take the average of the shifts of the O $1s$, Sr $3d$ and Nd $3d$ core levels as a measure of $\Delta \mu$ in NSMO \cite{Ebata}.
\begin{figure}
\begin{center}
\includegraphics[width=9cm]{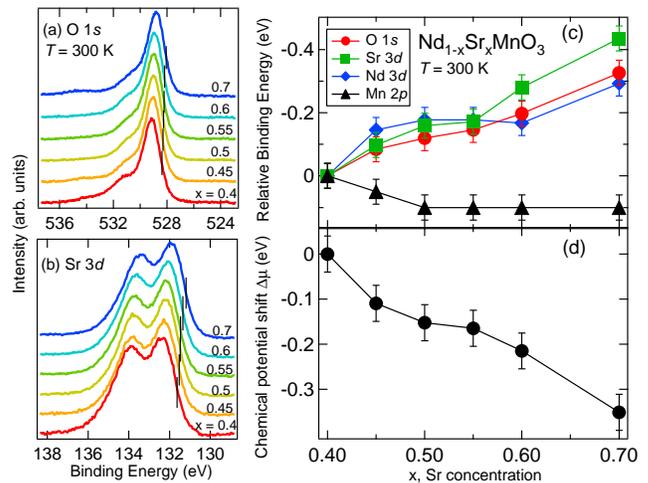}
\caption{(Color online) Core-level photoemission spectra and chemical potential shift in Nd$_{1-x}$Sr$_x$MnO$_3$ at 300 K. (a), (b) Photoemission spectra of the O $1s$ and Sr $3d$ core levels; (c) Binding energy shifts of the O $1s$, Sr $3d$, Nd $3d$, and Mn $2p$ core levels as functions of hole concentration $x$; (d) Chemical potential shift $\Delta \mu$ as a function of carrier concentration $x$.}
\label{chemicalpotential}
\end{center}
\end{figure}

In Fig. 2(d), we have plotted the $\Delta \mu$ of NSMO thus deduced as a function of carrier concentration. The measured shift is indeed due to the band structure and not caused by correlation effects of the high-energy photoemission process \cite{Ino, Satake}. A clear downward $\Delta \mu$ with hole concentration is observed in the regions $x \lesssim 0.45$ and $x \gtrsim 0.55$, as reported for LSMO \cite{Matsuno}. However, a weak but clear suppression of the shift is observed in and near the CE-type CO composition range $0.45 \lesssim x \lesssim 0.55$, corresponding to the narrow region of the CO phase in the phase diagram of NSMO as shown in Fig. 1 \cite{Kuwahara, Tokura5}. The PCMO shows a stronger suppression of $\Delta \mu$ in and near the wider CE-type CO composition range \cite{Ebata}. Such a chemical potential pinning was not observed in bulk and thin film LSMO \cite{Matsuno, Horiba} in which the CO phase does not exist and in PCMO thin films grown on LaAlO$_3$ substrates, where the CO phase was suppressed by compressive strain from the substrates, too \cite{Wadati2}. If the suppression is due to an electronic phase separation as the thermodynamic relationship suggests, the phase separation should occur only on a microscopic scale as described by the bi-stripe or Wigner-crystal model suggested for La$_{1-x}$Ca$_x$MnO$_3$ with $x \geq 0.5$ in order to avoid the accumulation of long-range Coulomb energy \cite{Mori2, Radaelli}. Although our measurements were performed for the PI phase above the CO transition temperature in NSMO, $\Delta \mu$ seemed to be influenced by the fluctuations of the CO state. This is consistent with the fact that such fluctuations have been observed in the PI phase of NSMO by means of x-ray scattering \cite{Kiryukhin}.

\begin{figure}
\begin{center}
\includegraphics[width=8.8cm]{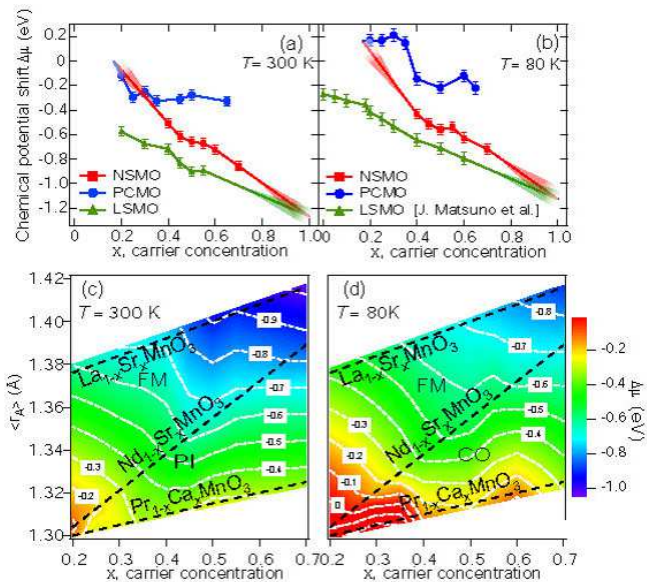}
\caption{(Color online) Chemical potential shift $\Delta \mu$ as a function of A-site average ionic radius $\langle r_A \rangle$ and hole concentration $x$. (a), (b) Comparison of the $\Delta \mu$ in Nd$_{1-x}$Sr$_x$MnO$_3$, Pr$_{1-x}$Ca$_x$MnO$_3$ and La$_{1-x}$Sr$_x$MnO$_3$ at 300 K and 80 K \cite{Ebata, Matsuno}. The relative chemical potential positions have been aligned so that the (extrapolated) data for the same $\langle r_A \rangle$ and $x$ coincide. (c), (d) $\Delta \mu$ interpolated from the data in (a) and (b) in the $x$-$\langle r_A \rangle$ plane at 300 K and 80 K \cite{Tokura5, Ebata, Matsuno}. FM, PI and CO denote approximate location of the ferromagnetic metallic, paramagnetic insulating and charge-ordering phases, respectively.}
\label{comparison}
\end{center}
\end{figure}

We have also determined the doping-dependent $\Delta \mu$ in LSMO at 300 K by measurements of core-level photoemission spectra as shown in Fig. 3(a). Here, we have deduced the doping-dependent $\Delta \mu$ of LSMO by taking the average of the shifts of the O $1s$, Sr $3d$ and La $3d$ core levels as before \cite{Matsuno}. The $\Delta \mu$ of LSMO at 300 K was almost the same as that at $\sim$ 80 K \cite{Matsuno}.  

In the rest of this paper, we shall deduce the chemical potential shift $\Delta \mu$ as a function of bandwidth, too.
First, we extrapolated each set of the doping-dependent $\Delta \mu$ data under the assumption that $\Delta \mu$'s for the same doping and the same A-site average ionic radius $\langle r_A \rangle$ should coincide, as shown in Fig. 3(a) and (b). For example, the PCMO and NSMO data at 300 K and 80 K have been aligned at $x=0.167$.
Here, the shift in PCMO at 80 K was strong in the CO region $x \lesssim 0.5$ compared to 300 K, as reported for valence-band spectra \cite{Ebata, Ebata2}. The incommensurate charge modulation in PCMO exists in a wider range of the hole concentration at 300 K than at 80 K as studied by electron diffraction or x-ray resonant scattering \cite{Milward}. Correspondingly, the suppression of $\Delta \mu$ in PCMO was observed in a wider region $x \gtrsim 0.3$ at 300 K than at 80 K in the CO region $x \gtrsim 0.5$. Thus the doping-dependent $\Delta \mu$ is well correlated with the change in the periodicity of stripes for PCMO \cite{Ebata2, Milward}.

\begin{figure}
\begin{center}
\includegraphics[width=12.8cm]{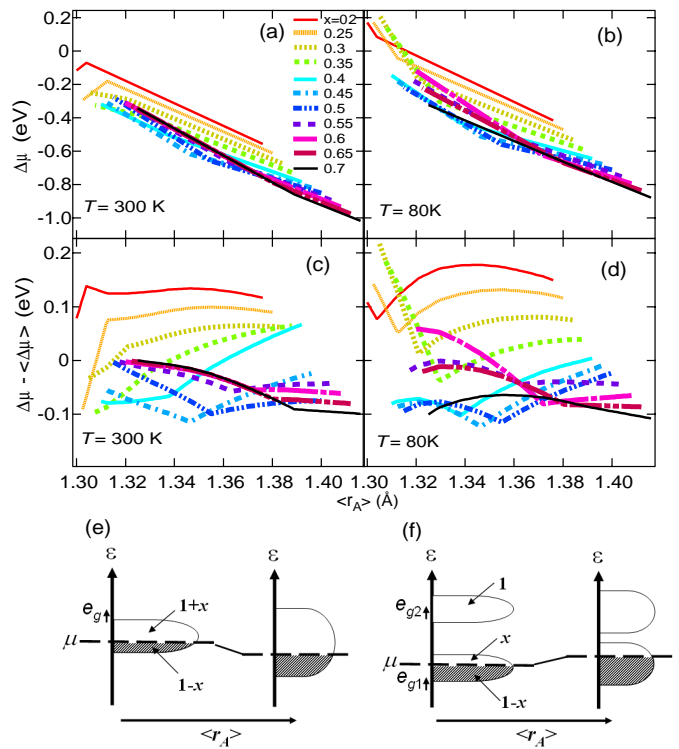}
\caption{(Color online) Chemical potential shift $\Delta \mu$ as a function of the A-site ionic radius $\langle r_A \rangle$. (a) $\Delta \mu$ at 300 K; (b) $\Delta \mu$ at 80 K; (c) Difference between the $\Delta \mu$ and the average shift $\Delta \mu - \langle \Delta \mu \rangle$ at 300 K; (d) $\Delta \mu - \langle \Delta \mu \rangle$ at 80 K; (e) Schematic pictures of the density of states (DOS) in the region of ferromagnetic metallic (FM) phase $x<0.5$ for degenerate two-orbital model; (f) Schematic pictures of the DOS in the region of FM phase $x<0.5$ for one-orbital model resulting from the Jahn-Teller splitting.}
\label{rA}
\end{center}
\end{figure}

Figure 3(c) and (d) shows $\Delta \mu$ in the $x$-$\langle r_A \rangle$ plane and the resulting $\Delta \mu$ is linearly interpolated from the experimental data \cite{Tokura5, Ebata, Matsuno}. In large-bandwidth systems without CE-type CO region, the chemical potential is shifted monotonously with carrier concentration because stripe effects are negligibly weak. In the intermediate- and small-bandwidth regions, $\Delta \mu$ curves show characteristic doping dependences in the sense that the pinning of chemical potential as a function of $x$ due to the incommensurate charge modulation was observed in and near the CE-type CO composition range. The clear contrast in the doping dependence of the chemical potential shifts between the wide bandwidth and narrow bandwidth systems reflects the existence and absence of the CE-type CO phase, respectively. The results thus indicate again that the tendency toward charge self-organization around $x$ $\sim$ 0.5 such as stripe formation is indeed enhanced with decreasing bandwidth.

If the chemical potential shift $\Delta \mu$ is plotted as a function of $\langle r_A \rangle$ as shown in Fig. 4(a) and (b), a downward $\Delta \mu$ with increasing $\langle r_A \rangle$ is commonly observed. The overall downward shift with $\langle r_A \rangle$ would be related to the decrease of the orthorhombic distortion.
Because the hybridization between the $e_g$ and $t_{2g}$ orbitals caused by the orthorhombic distortion raises the $e_g$ level and lowers the $t_{2g}$ level, the weakening of the $e_g$-$t_{2g}$ hybridization will lower the $e_g$ level and thus lowers the chemical potential.
The large $\langle r_A \rangle$'s on large-bandwidth systems are known as conducting ferromagnets, where the itinerant $e_g$ electrons mediate ferromagnetic interaction between neighboring Mn$^{3+}$ and Mn$^{4+}$ ions through double-exchange interaction. In order to extract $\Delta \mu$ induced by double-exchange interaction, we have subtracted the doping averaged chemical potential shift $\langle \Delta \mu \rangle$ as a function of $\langle r_A \rangle$ from the measured $\Delta \mu$. Here, we consider that the shift due to the orthorhombic distortion does not depend on the hole doping. In Fig. 4(c) and (d), we have plotted the resulting differences $\Delta \mu - \langle \Delta \mu \rangle$ as a function of $\langle r_A \rangle$. In the composition range $0.3<x<0.5$ where the FM phase appears, one can see an upward shift $\Delta \mu - \langle \Delta \mu \rangle$ with increasing $\langle r_A \rangle$ and hence increasing bandwidth. The upward $\Delta \mu$ with decreasing temperature has been observed in the low-temperature region of the FM phase of NSMO with $x=0.4$ and 0.45 because of double-exchange interaction with the Jahn-Teller-split $e_g$ band \cite{Ebata3}. Furukawa \cite{Furukawa} has predicted theoretically using the one-orbital model that in a double-exchange system a large upward $\Delta \mu$ occurs with decreasing temperature in the FM phase due to the increase of the bandwidth. If the $e_g$ band remains doubly degenerate in the FM phase for $x<0.5$, one would expect to see a downward $\Delta \mu$ with increasing bandwidth [see, Fig. 4(e)]. Therefore, we attribute the upward shifts with increasing bandwidth to the double-exchange interaction with the Jahn-Teller-split $e_g$ band as shown in Fig. 4(f), consistent with the temperature-dependent $\Delta \mu$ \cite{Ebata3}.

For the bandwidth $\langle r_A \rangle$-controlled chemical potential shift, if ${\partial\mu}/{\partial \langle r_A \rangle } \sim 0$, there is a possibility of separation into two ``phases" with different $\langle r_A \rangle$ values with the same $x$. In the bandwidth-controlled system Pr$_{0.55}$(Ca$_{1-y}$Sr$_y$)$_{0.45}$MnO$_3$, it is reported that there is a bicritical point in the electronic phase diagram at around $y \sim 0.25$, in which the CE-type CO and the FM phases compete each other \cite{Tomioka_PCSMO}. Since the present results as a function of $\langle r_A \rangle$ have only a few data points for each $x$, more detailed measurements are necessary to explore whether ${\partial\mu}/{\partial \langle r_A \rangle } \sim 0$ occurs near the FM-CO boundary in the manganite phase diagram.

In conclusion, we have experimentally determined the doping and bandwidth dependences of the chemical potential in the perovskite-type manganites by means of core-level photoemission measurements of LSMO, NSMO, and PCMO. We observed a suppression of the $\Delta \mu$ for hole concentration near and in the CE-type CO composition range. We have found correlation between this suppression and the change of the periodicity of stripes with hole doping. Also, we found a downward $\Delta \mu$ with increasing $\langle r_A \rangle$ due to the decrease of the orthorhombic distortion. After subtracting the filling independent part of this shift, an upward $\Delta \mu$ with increasing bandwidth was found to be realized in the FM phase for $0.3<x<0.5$, which we attribute to the double-exchange interaction in the Jahn-Teller-split $e_g$ band.

Experimental support by H. Wadati, K. Maekawa, T. Kataoka and K. Ishigami is gratefully acknowledged. This work was supported by a Grant-in-Aid for Scientific Research in Priority Area ``Invention of Anomalous Quantum Materials" from the Ministry of Education, Culture, Sports, Science and Technology, Japan.

\end{document}